\documentclass[preprint,aps,superscriptaddress,nofootinbib]{revtex4}

\usepackage{amsmath}
\usepackage{graphicx}
\usepackage{color}
\usepackage{slashed}

\begin{document}

\title{Black hole remnant in asymptotic Anti-de Sitter space}
\author{Wen-Yu Wen}\thanks{%
E-mail: steve.wen@gmail.com}
\affiliation{Department of Physics and Center for High Energy Physics, Chung Yuan Christian University, Chung Li City, Taiwan}
\affiliation{Leung Center for Cosmology and Particle Astrophysics\\
National Taiwan University, Taipei 106, Taiwan}
\author{Shang-Yu Wu}\thanks{%
E-mail: loganwu@gmail.com}
\affiliation{Department of Electrophysics, National Chiao Tung University, Hsinchu, Taiwan}

\begin{abstract}
It is known that a solution of remnant were suggested for black hole ground state after surface gravity is corrected by loop quantum effect.  On the other hand, a Schwarzschild black hole in asymptotic Anti-de Sitter space would tunnel into the thermal soliton solution known as the Hawking-Page phase transition.   In this letter, we investigate  the low temperature phase of three-dimensional BTZ black hole and four-dimensional AdS Schwarzschild black hole.  We find that the thermal soliton is energetically favored than the remnant solution at low temperature in three dimensions, while Planck-size remnant is still possible in four dimensions.   Though the BTZ remnant seems energetically disfavored, we argue that it is still possible to be found in the overcooled phase if strings were present and its implication is discussed.
\end{abstract}

\maketitle



\section{Introduction}

It was shown that for a Schwarzschild black hole evaporating massless particles, the Hawking-Bekenstein area law \cite {Bekenstein:1973ur} receives the logarithmic correction, $S = 4\pi M^2 - 4\pi\sigma \log (M+\sigma)$ for one-loop corrected temperature $T=(8\pi M)^{-1}(1+\sigma( M^2)^{-1} )$, where $\sigma$ is the conformal anomaly.  In particular, a Planck size remnant is implied for $\sigma<0$ \cite{Fursaev:1994te}.  There were complaints about remnants such as one in the \cite{Susskind:1995da}, mainly due to extraordinary large amount of entropy confined within a tiny volume.  Nevertheless, if the Hawking radiation were not exactly thermal, the retained information would have been released through the evaporation process.  One could have expected the existence of a remnant with zero or small amount of residual entropy.  In this case, instead to help resolve the information loss paradox, the remnant solution might still answer the call for the hypothetically fundamental Planck scale.  In general, remnant solutions can be easily found in those theories such as noncommutative geometry, doubly special relativity, or generalized uncertainty principle (GUP).  Readers are directed to a recent review \cite{Chen:2014jwq} for instance.

On the other hand, it is well known that a Schwarzschild black hole in asymptotic Anti-de Sitter space would tunnel into the thermal AdS solution known as the Hawking-Page phase transition\cite{Horowitz:1998ha,Surya:2001vj}.  In particular, the three dimensional Banados-Teitelboim-Zanelli (BTZ) black hole system \cite{Banados:1992wn} has two distinct phases: the black hole with mass $M>0$ and thermal AdS with $M=-1$ \cite{Eune:2013qs}.  It is unclear for us whether this thermal AdS phase is still energetically favored at low temperature after the BTZ or AdS black hole receives the loop quantum correction.  In other words, there might be a possibility that the black hole would have already stopped evaporating and stayed as a warm remnant before it enters the soliton phase.

To proceed our discussion, one needs to assume that both first law of thermodynamics and logarithmic correction to the black hole entropy are valid within the energy range in our discussion.  The extrapolation of both relations to the limit of Planck size maybe too naive, especially for our ignorance of complete theory of quantum gravity.  However, our strategy is to compare free energy of remnant and that of thermal AdS around the Hawking-Page temperature $T_{HP}=1/2\pi\l$.  This corresponds to the energy scale ${\cal O}(\l^{-1})$, which is still far from the Planck or string scale, it should be reasonable to assume the laws of thermodynamics are valid and the black hole size around ${\cal O}(\l)$ can be treated as classical and static background.

At last, we comment that the logarithmic correction remains a simialr form as long as in the dual CFT side, the number of states reads\cite{Carlip:2000nv}
\begin{equation}
\rho \sim c^{\gamma}e^{S},
\end{equation}
where $c$ is the central charge, $\gamma$ is some rational power, and $S$ is the entropy to be identified with the Bekenstein-Hawking entropy in the black hole side.  We do not expect this Cardy formula to be abruptly modified for large or small BTZ black hole.  Another supportive though hand-waving argument comes from the GUP modified Schwarzschild black hole entropy\cite{Adler:2001vs}:
\begin{equation}
S_{GUP} = 4\pi f(M^2)M^2-log(M+\sqrt{M^2-1}),
\end{equation}
where black hole mass $M$ is measured in the unit of Planck mass and function $f(M^2)$ interpolates between $0$ and $1$ for $0\le M<\infty$.  It is obvious the logarithmically corrected form survives for all $M>1$.

This letter is organized as follows.  We review the Hawking-Page phase transition between BTZ black hole and thermal AdS$_3$ in the section II.  In the section III, we compute up to two loop-corrected entropy and free energy for BTZ black hole and discuss its phase transition to thermal AdS.  In the section IV, we calculate the one-loop corrected entropy and free energy for AdS$_4$ black hole and discuss tis phase transition.  In the section V, we discuss a new scale set by the AdS remnant and its effect on the Hawking-Page phase transition.  At last, we discuss possible scenarios  around phase transition if stringy excitation is considered.

\section{BTZ black hole and Hawking-Page phase transition}

To justify which phase is energetically favored at specific temperature, we would like to compare free energy of BTZ and that of vacuum.  Since we expect a vacuum without nonzero angular momentum,  we will only consider nonrotating BTZ black hole.  With that being said, we will not inspect solutions such as extremal BTZ solution or exotic BTZ\cite{Townsend:2013ela}.  Although the phase diagram would become rich and interesting by including those solutions, they cannot tunnel into AdS vacuum as long as conservation of angular momentum is respected.  The nonrotating BTZ solution has the metric
\begin{equation}
ds^{2}=-(-M+\frac{r^2}{\l^2})dt^2+(-M+\frac{r^2}{\l^2})^{-1}dr^2+r^2 d\phi^2
\end{equation}
with horizon $r_+=\l \sqrt{M}$.  The thermal quantities, such as Hawking temperature $T_{H}$, Bekenstein-Hawking entropy $S_{BH}$, internal energy $E$ and free energy $F_{BH}$, are given by
\begin{eqnarray}
&&T_{H}=\frac{\kappa}{2\pi}=\frac{\sqrt{M}}{2\pi \l},\\
&&S_{BH} = \frac{A_H}{4G}=\frac{\pi \l \sqrt{M}}{2G},\\
&&E=\frac{M}{8G},\\
&&F_{BH}= E-T_H S_{BH} = -\frac{M}{8G}.
\end{eqnarray}

There also exists a thermal AdS soution with on shell action $I$ and free energy $F_{AdS}$:

\begin{eqnarray}
&&I = -\frac{\beta}{8G},\nonumber\\
&&F_{AdS} = -\frac{1}{8G}
\end{eqnarray}
and the phase transition occurs at temperature $T=1/(2\pi\l)$, when the free energy of nonrotating BTZ becomes higher than that of thermal vacuum.

\section{Quantum correction to surface gravity and the emergence of the remnant}
The quantum correction to the Hawking temperature has following form\cite{Liu:2014}
\begin{equation}\label{temp_corrected}
T^{q}_{H} = T_{H} /(1+\sum_i\alpha_i\frac{\hbar^i}{r_+^i}).
\end{equation}
In generic, the coefficients $\alpha_i$ will depend on species of particles included in the $i$-th loop perturbative correction\cite{Fursaev:1994te}.  Following the first law of thermodynamics, the entropy receives corresponding correction as follows:
\begin{equation}\label{entropy_corrected}
S^{q\prime}_{BH} = \int{\frac{dM}{T^q_{H}}}=\frac{\pi r_+}{2G} + \alpha_1\hbar\frac{\pi}{2G}\ln r_+ - \alpha_2 \hbar^2 \frac{\pi}{2G r_+} +-\cdots
\end{equation}
The conformal field theory (CFT) calculation also implies a similar logarithmic correction with coefficient $\alpha_1=-\frac{3G}{\pi\hbar}$ in the above expression\cite{Carlip:2000nv}.  However, we remark that the temperature in the CFT computation was not modified according to (\ref{temp_corrected}).  Therefore, it is more appropriate to think the logarithmic correction we will discuss here comes from a different origin and therefore $\alpha_1$ could be negative but varied for different models of quantum gravity.  Nevertheless, the expression (\ref{entropy_corrected}) is not UV complete for it suffers from the divergence as $M\to 0$.  Without knowing much about gravity at the Planck scale, we would like to assume the entropy takes the same expression as (\ref{entropy_corrected}) for simplicity.   Therefore, one can regard the undetermined integral constants as counterterms such that expression (\ref{entropy_corrected}) can be regularized in each loop computation.  Moreover, the uniqueness of ground state implies that the would-be regularized $S^q_{BH}$ must vanish at some critical mass $m_c$, which can be regarded as the remnant mass.  At last, the finite expression suggests that
\begin{equation}
S^{q}_{BH} =\frac{\pi \l \sqrt{M}}{2G} -\frac{\pi \l \sqrt{m_c}}{2G} + \alpha_1\hbar\frac{\pi}{4G}\ln (M/m_c) - \alpha_2 \hbar^2 \frac{\pi}{2G\l\sqrt{M}}+ \alpha_2 \hbar^2 \frac{\pi}{2G\l\sqrt{m_c}} +-\cdots
\end{equation}

\begin{figure}[tbp]
\includegraphics[width=0.6\textwidth]{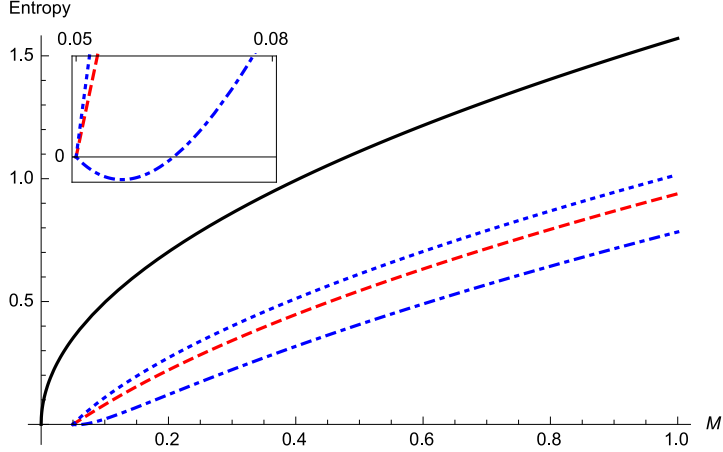}
\caption{\label{fig:entropy} A numerical simulation of entropy for BTZ blakc hole (black solid), one-loop corrected remnant(red dash), two-loop corrected remnant with $\alpha_2>0$ (blue dot) and two-loop corrected remnant with $\alpha_2<0$ (blue dot-dash).  The inset zoom-in graph shows some exotic behavior near the critical remnant mass for $\alpha_2<0$.   This unphysical region of negative entropy should be excluded and remnant mass is chosen to be the larger root to equation $S^{q}_{BH}=0$.}
\end{figure}

In the Fig.\ref{fig:entropy}, we plot the entropy versus mass for BTZ black hole and its quantum correction.  Though there is no constraint on the coefficient of second loop $\alpha_2$, the numerical result indicates that there is a small region with negative entropy for the case $\alpha_2<0$.   This unphysical resion should be excluded and the remnant mass is redefined as the larger positive root to equation $S^{q}_{BH}=0$ in the case $\alpha_2<0$.

\begin{figure}
\includegraphics[width=0.6\textwidth]{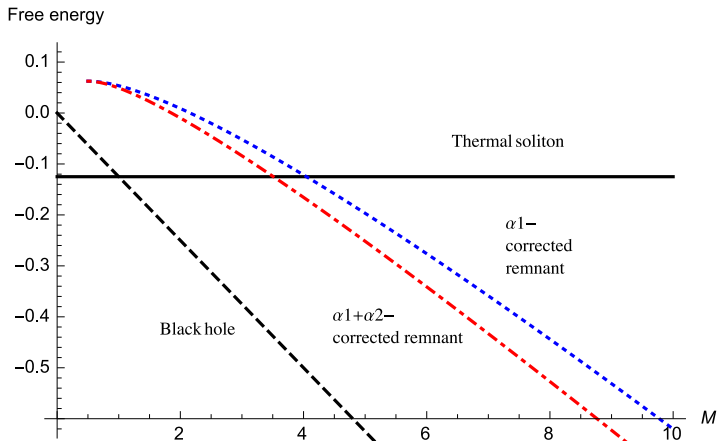}
\caption{\label{fig:free_energy} A numerical simulation of free energy for BTZ blakc hole (black dash), one-loop corrected remnant (blue dot), two-loop corrected remnant (red dot-dash) and thermal soliton (black solid).}
\end{figure}

At last, the quantum corrected free energy can be computed as
\begin{equation}
F^q_{BH} = E - T^q_{H} S^{q}_{BH}
\end{equation}
We simulate the free energy of quantum-corrected solutions and plot them in the Fig,\ref{fig:free_energy}.  We find that with quantum correction, the Hawking-Page transition happens at slightly larger $M$ and the black hole turns into the AdS thermal vacuum before a remnant could possibly form.   This result can be partially understood by absorbing those quantum corrections in (\ref{temp_corrected}) into a definition of {\sl effective} horizon size $r_+^{eff} \equiv r_+ - \alpha_1\hbar - \cdots$ for $r_+ \gg \l_p$.  That is, the quantum fluctuation makes the horizon look smaller than its classical value, therefore less stable (for small black hole).  However, unless the quantum corrected black hole stop radiation before it hits the Hawking-Page temperature, it eventually becomes energetically disfavored and decayed into the thermal vacuum.  This is what happen to the BTZ case.  In the following, we will examine phase transition of the Schwarzschild black hole in AdS$_4$.

\section{Schwarzschild-AdS$_4$ black hole}
In this section, we would like to compute the free energy of Schawarzschild black hole in AdS$_4$.  Given the black hole metrics
\begin{eqnarray}
&&ds^2 = -V(r) dt^2 + V(r)^{-1}dr^2 + r^2 d\Omega_2^2,\nonumber\\
&&V(r) = 1-\frac{2GM}{r}+\frac{r^2}{\l^2}
\end{eqnarray}
The thermal quantities read
\begin{eqnarray}
&&T_H=\frac{\l^2+3r_+^2}{4\pi \l^2 r_+} \\
&&S_{BH}=\pi r_+^2\\
&&F_{BH}=-\frac{r_+(r_+^2-\l^2)}{4\l^2}
\end{eqnarray}
where $V(r_+)=0$ is satisfied.  The free energy is defined with respect to the  AdS soliton, that is $F_{AdS}=0$.  The Hawking-Page phase transition happens at $T_{HP}=1/(\pi\l)$, where $r_+=\l$.

The loop corrected thermal quantities after regularization are given by\cite{Banerjee:2008ry}

\begin{eqnarray}
&&T_H^q = T_H (1+\frac{\alpha}{M^2})^{-1},\\
&&S_{BH}^q=\int{\frac{dM}{T_H^q}}=S_{BH}-\pi r_c^2 +4\pi \alpha \frac{\l^2(r_c^2-r_+^2)}{(\l^2+r_+^2)(\l^2+r_c^2)} -4\pi \alpha \ln\frac{r_c^2(r_+^2+\l^2)}{r_+^2(r_c^2+\l^2)},\\
&&F_{BH}^q=E -T^q_H S_{BH}^q
\end{eqnarray}

\begin{figure}[tbp]
\includegraphics[width=0.6\textwidth]{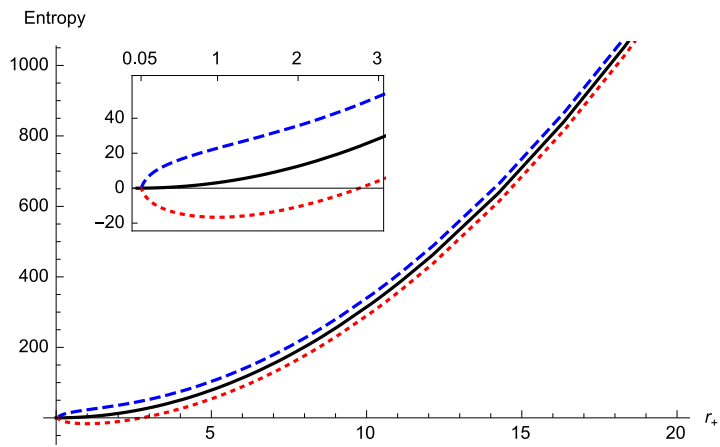}
\caption{\label{fig:entropy2} A numerical simulation of entropy for AdS blakc hole (black solid), one-loop corrected remnant(blue dash) with $\alpha<0$, and one-loop corrected remnant with $\alpha>0$ (red dot).  The inset zoom-in graph shows some exotic behavior for $\alpha>0$ near the remnant crtitcal mass, where the unphysical region of negative entropy should be excluded.  We has used the unit $\l=1$ in the plot.}
\end{figure}

In the Fig.\ref{fig:entropy2}, we plot the entropy versus horizon radius for AdS black hole and its quantum correction.  The unphysical region of negative entropy should be excluded in the case $\alpha>0$ and the remnant mass is chosen to be the larger root to the equation $S_{BH}^q=0$ .  On the other hand, the case of negative $\alpha$ is consistent with previous observation from AdS-CFT correspondence\cite{Carlip:2000nv} and argument based on remnant\cite{Xiang:2006mg,Chen:2009ut}.

\begin{figure}
\includegraphics[width=0.6\textwidth]{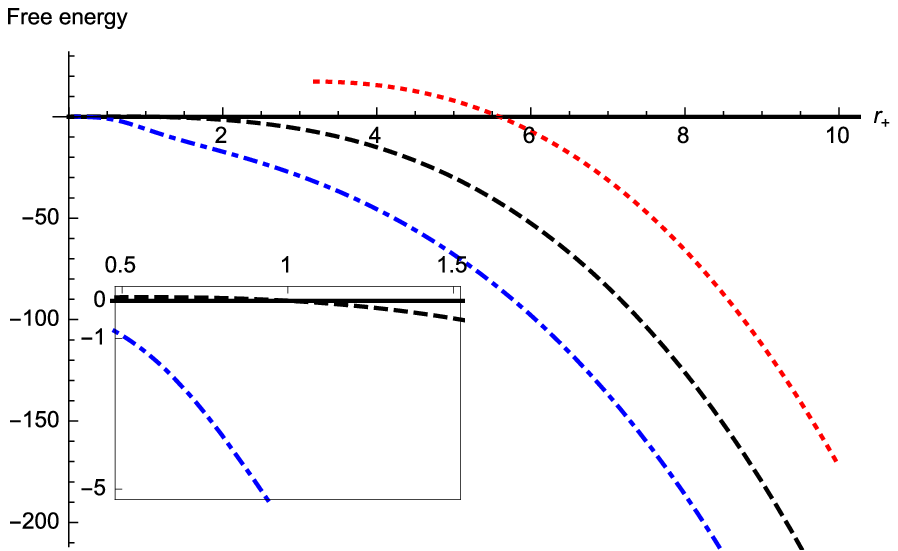}
\caption{\label{fig:free_energy2} A numerical simulation of free energy for AdS blakc hole (black dash),  one-loop corrected remnant(blue dash) with $\alpha<0$, one-loop corrected remnant with $\alpha>0$ (red dot) and thermal soliton (black solid or x-axis).  The inset zoom-in graph shows that AdS soliton is energetic favored than black hole below the critical size $r_+=\l$, nevertheless a one-loop corrected remnant phase is favored for $\alpha<0$.  We has used the unit $\l=1$ in the plot.}
\end{figure}

In the Fig.\ref{fig:free_energy2}, we plot the free energy versus horizon radius for AdS black hole and its quantum correction.  It shows that below some critical size, thermal AdS soliton is energetic favored than AdS black hole and its one-loop corrected remnant with positive $\alpha$.  Nevertheless the one-loop corrected remnant with negative $\alpha$ seems to survive all the way to the Planck size.

\section{Implication of AdS remnant}

The partition functions of thermal AdS$_3$ and BTZ black hole are related by a modular transformation\cite{Hubeny:2009rc}
\begin{equation}\label{partition}
Z_{BTZ}(T_{BTZ},\Omega) = Z_{AdS}(T_{AdS},\Omega),\qquad T_{AdS}=\frac{1-\Omega^2\l^2}{4\pi^2 T_{BTZ}\l^2}
\end{equation}
where $\Omega=\frac{r_-}{r_+\l}$ is the chemical potential conjugated to the angular momentum of BTZ.  In the nonrotating case, the Hawking-Page transition temperature is defined as $T_{AdS}=T_{BTZ}=(2\pi\l)^{-1}$.  Since the loop correction does not modify the metric, this discrete symmetry is believed to persist.  Therefore, the remnant temperature implies a cutoff temperature in thermal AdS phase, that is
\begin{equation}
T^c_{AdS} = \frac{1}{2\pi\l \sqrt{m_c}}
\end{equation}
A geometric picture is available to explain this relation\cite{Maldacena:1998bw} that one can view the Euclidean BTZ or thermal AdS as a hyperbolic three-manifolds with a two-torus as its boundary.  The choice of one cycle gives the description of temperature in either BTZ or thermal AdS$_3$ and the modular transformation (\ref{partition}) simply swaps the cycle.  While the loop correction to Hawking temperature obstructs the unlimited expansion of Euclidean time-cycle of BTZ, it also creates a lower bound such that the period of Euclidean time-cycle of AdS cannot be less than $1/T^c_{AdS}$.

On the other hand, if the black hole stops radiation at the temperature of remnant, it also implies a minimum length scale $\tau_2^c$ associating to the remnant size, such that the periodicity of cycle $\tau_2 \ge \tau_2^{c}$.  Using the same method to determine the Hagedorn temperature $\beta_H^{-1}$, now the partition function for lowest excitation modes becomes\cite{Berkooz:2007fe}
\begin{equation}\label{hagedorn}
Z \sim e^{-\frac{1}{4\pi\tau_2^c}(\beta^2-\beta_H^2)}.
\end{equation}
The available states above the Hagedorn temperature ($\beta<\beta_H$) still grow exponentially but may not be as sharp as that of usual BTZ where the point-like string limit $\tau_2\to 0$ can be taken.  That is, the existence of BTZ remnant may also smooth out the phase transition for its slower growth of partition function.

\section{Stringy point of view}
\begin{figure}[tbp]
\includegraphics[width=0.6\textwidth]{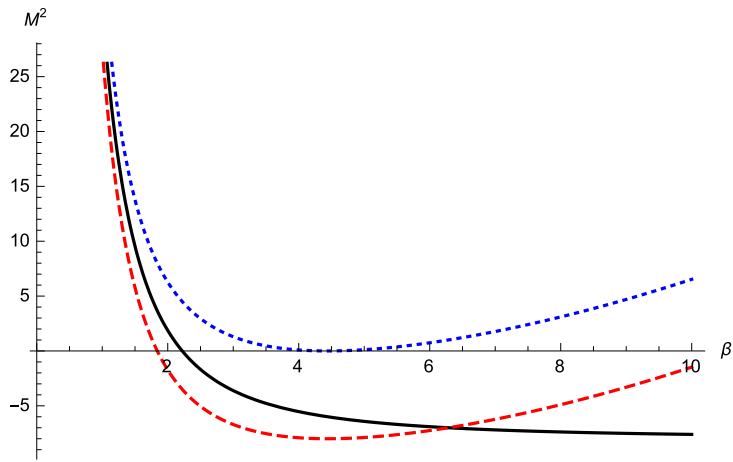}
\caption{\label{fig:Atik} Stringy states of pure momentum mode $(m,n)=(\pm1,0)$ (Black solid) become tachyonic for $M^2<0$ at small temperature.  They can be stabilized if nonzero winding modes (for non-contractable Euclidean time cycle of remnant) are also present.  We plot states of $(\pm1,\pm1)$ (Blue dotted), and $(\pm 1,\mp 1)$ (Red dashed) }.
\end{figure}

Earlier it has been shown that BTZ remnant (up to two-loop correction) is not energetically favored below the critical temperature, nevertheless we would like to argue that it may still survive as a metastable state during overcooling phase.  First we have learnt that if there is stringy excitation, there would appear two Hagedorn temperatures $T_{H/L}$, which are different from the Hawking-Page transition temperature by\cite{Berkooz:2007fe}

\begin{equation}
T_{L}=T_{HP}\frac{1}{\sqrt{k}}(4-\frac{1}{k-2})^{1/2}=\frac{4\pi^2\l^2}{T_{H}}, \qquad T_{HP}=\frac{1}{2\pi}
\end{equation}
where strings propagating in the AdS$_3$ space are described by the $SL(2,R)$ WZW model at level $k$.  The BTZ could be overcooled between $T_L$ and $T_{HP}$, while thermal AdS could be overheated between $T_{HP}$ and $T_H$.   If the critical remnant mass locates in the following range

\begin{equation}
1>m_c>\frac{1}{k}(4-\frac{1}{k-2})
\end{equation}
then the remnant is still likely to be observed as a metastable state in this overcooled phase.  To further illustrate this possibility, we recall that while the higher Hagedorn temperature $T_H$ can be understood as appearance of  the Atick-Witten tachyon winding mode\cite{Atick:1988si},  the lower temperature $T_L$ can also be understood as appearance of the tachyonic momentum mode as shown in the Figure \ref{fig:Atik}.  The previous-mentioned obstruction of unlimited expansion of Euclidean time-cycle of BTZ can be understood as appearance of a winding mode to stabilize this tachyon.  To illustrate this point, let us consider a toy model of string with compactified time dimension of circumference $\beta$, that is
\begin{equation}
X^0(\sigma,\tau)=x^0 + \frac{2\pi m \tau}{\beta}+\frac{n\beta\sigma}{\pi}+\cdots,
\end{equation}
with mass shell condition
\begin{eqnarray}
\frac{M^2}{4} =&& N+\frac{1}{2}(\frac{m\pi}{\beta}+\frac{n\beta}{2\pi})^2-1 \nonumber\\
&&+ \tilde{N} +\frac{1}{2}(\frac{m\pi}{\beta}-\frac{n\beta}{2\pi})^2-1
\end{eqnarray}
satisfying level matching constraint $N-\tilde{N}=mn$.  In the Figure \ref{fig:Atik}, we show that for pure momentum mode $(m,n)=(\pm1,0)$, the state becomes tachyonic for large enough $\beta$(such that temperature lower than $T_L$), however, this tachyonic state could be stabilized if at the same time a winding mode associated with remnant is present, say $(m,n)=(\pm 1,\pm1)$.  This metastable remnant solution could be understood from the partition function (\ref{hagedorn}) as well.  If the scale $\tau_2^c \sim {\cal O}(1/T_L)$, then the phase transition would be too mild to happen, or the difference between two phases is hard to distinguish.  In the former case, the BTZ phase remains (as a remnant); while in the latter, it could be a coexistent phase for both thermal AdS and BTZ.

\section{Summary}

In this letter, we investigated the low temperature phase of three-dimensional BTZ black hole and four-dimensional AdS Schwarzschild black hole.  We found that the thermal AdS is energetically favored than the remnant solution at low temperature in three dimensions, while Planck-size remnant is still possible in four dimensions for negative one-loop coefficient.   Though the BTZ remnant seems energetically disfavored, we argue that it is still possible to be found in the overcooled phase if strings were present.  At last, we showed that existence of BTZ remnant scale might have effect to smooth the change of degrees of freedom during Hawking-Page transition.   In order to justify our conclusions derived from black hole thermodynamics and loop-corrected Hawking temperature, one may need to sutdy possible candidates for microstates which are responsible to the thermal properties of BTZ remnant.  In other words, one would ask that after introducing loop quantum correction, whether a new mechanism for phase transition between BTZ and remnant could exist in the dual CFT description.  We will leave this for future study.

\begin{acknowledgments}
WYW is grateful to the hospitality of YITP and LeCosPA for fruitful discussion with Hirotaka Irie and Pisin Chen during the early stage of this project.  WYW is also grateful to the hospitality of Osaka University for useful feedbacks from Koji Hashimoto, Norihiro Iizuka, Satoshi Yamaguchi and Ohta Nobuyoshi.  This work is supported in parts by the Taiwan's Ministry of Science and Technology (grant No. 102-2112-M-033-003-MY4) and the National Center for Theoretical Science. SYW was supported by the Ministry of Science and Technology (grant No. MOST-101-2112-M-009-005 and MOST 104-2811-M-009-068) and National Center for Theoretical Science in Taiwan.
\end{acknowledgments}


\end{document}